\documentclass[10pt,letterpaper]{IEEEtran}

\usepackage{amsmath}
\usepackage{amssymb}
\usepackage{epsfig}
\usepackage{tabularx}
\usepackage{stfloats}
\usepackage{url}
\usepackage{balance}
\usepackage[noadjust]{cite}

\newtheorem{definition}{Definition}[section]
\newtheorem{lemma}{Lemma}[section]
\newtheorem{example}{Example}[section]

\makeatletter
\def\keywords{\normalfont%
    \if@twocolumn%
      \@IEEEabskeysecsize\bfseries\textit{Keywords}---\,\relax%
    \else%
      \begin{center}\@IEEEabskeysecsize\bfseries Keywords\end{center}\quotation\@IEEEabskeysecsize%
    \fi\@IEEEgobbleleadPARNLSP}
\def\endkeywords{\relax\if@technote\vspace{1.34ex}\else\vspace{0.67ex}\fi%
    \par\if@twocolumn\else\endquotation\fi%
    \normalsize\normalfont}
\makeatother

\title{\boldmath GF($2^m$) Low-Density Parity-Check Codes Derived from Cyclotomic Cosets \unboldmath}

\author{
\authorblockN{C. Tjhai, M. Tomlinson, R. Horan, M. Ambroze and M. Ahmed\\}
\authorblockA{Fixed and Mobile Communications Research,\\University of Plymouth,\\Plymouth PL4 8AA,\\United Kingdom,\\
             email: \{ctjhai,mtomlinson,rhoran,mambroze,mahmed\}@plymouth.ac.uk}
}

\begin{document}

\maketitle

\begin{abstract}
\boldmath
Based on the ideas of cyclotomic cosets, idempotents and Mattson-Solomon polynomials, we present a new method to construct
GF($2^m$), where $m>0$ cyclic low-density parity-check codes. The construction method produces the dual code idempotent
which is used to define the parity-check matrix of the low-density parity-check code. An interesting feature of this
construction method is the ability to increment the code dimension by adding more idempotents and so steadily decrease
the sparseness of the parity-check matrix. We show that the constructed codes can achieve performance very close to the
sphere-packing-bound constrained for binary transmission.
\unboldmath
\end{abstract}


\section{Introduction}\label{sec:introduction}
Since the recent rediscovery of low-density parity-check (LDPC) codes, a great deal of effort has been devoted to
constructing LDPC codes that can work well with the belief-propagation iterative decoder.
The studies of long block-length LDPC codes are very much established. The recent works of~\cite{Richardson_et_al.2001b},
\cite{Chung_et_al.2001b} have shown that, for long block-lengths, the best performing LDPC codes are irregular codes
and these codes can outperform turbo codes of the same block-length and code-rate. These long LDPC codes have degree
distributions which are derived from differential evolution~\cite{Richardson_et_al.2001b}
or Gaussian Approximation~\cite{Chung_et_al.2001a}. It can be shown that, using the concentration
theorem~\cite{Richardson_et_al.2001a}, the performance of infinitely long LDPC codes of a given degree distribution can
be characterised by the average performance of the ensemble based on cycle-free assumption. This assumption, however,
does not work for short and moderate block-length LDPC codes due to the inevitable existence of cycles in the underlying
Tanner Graphs. Consequently, for a given degree distribution, the performance of short block-length LDPC codes varies
considerably from the ensemble performance. Various methods exist for the construction of finite block-length irregular
codes~\cite{Campello_et_al.2001},\cite{Hu_et_al.2002},\cite{Ramamoorthy_et_al.2004b}. In addition to irregular LDPC
codes, algebraic constructions exist and the resulting codes are regular and usually cyclic in nature. Some examples
of algebraic LDPC codes are the Euclidean and Projective Geometry codes~\cite{Kou_et_al.2001}.

It has been noticed by the authors that, in general, there is a performance association between the code minimum distance
($d_{min}$) and decoding convergence. The irregular LDPC codes converge very well with iterative decoding, but their
$d_{min}$ are reasonably low. On the other hand, the algebraically constructed LDPC codes, which have high $d_{min}$,
tend not to converge well with the iterative decoder. It is not surprising that algebraically constructed codes may
outperform the irregular codes. The latter have error-floor which is caused by the $d_{min}$ error-events.
On the encoding side, the existence of algebraic structure in the codes is of benefit. Rather than depending on the
parity-check or generator matrices for encoding, as in the case of irregular codes, a low-complexity encoder 
can be built for the algebraic LDPC codes. One such example is the linear shift-register encoder for cyclic LDPC codes.
Assuming that $n$ and $k$ denote the codeword and
information length respectively, algebraic codes that are cyclic offer another decoding advantage. The
iterative decoder has $n$ parity-check equations to iterate with instead of $n-k$ equations, as in the case of
non-cyclic LDPC codes, and this leads to improved performance.

It has been shown that the performance of LDPC codes can be improved by going beyond the binary field~\cite{Davey_et_al.1998},
\cite{Hu_et_al.2005}. Hu \textit{et al.} showed that, under iterative decoding, the non binary LDPC codes have better
convergence properties than the binary codes~\cite{Hu_et_al.2005}. They also demonstrated that a coding gain of $0.25$dB
is achieved by moving from $\text{GF}(2)$ to $\text{GF}(2^6)$. Non binary LDPC codes in which each symbol takes values
from GF($2^m$) offer an attractive scheme for higher-order modulation. The complexity of the symbol-based iterative decoder
can be simplified as the extrinsic information from the component codes can be evaluated using the frequency domain dual
codes decoder based on the Fast-Walsh-Hadamard transform.

Based on the pioneering works of MacWilliams~\cite{MacWilliams_et_al.1977},\cite{MacWilliams.1979} on the
idempotents and the Mattson-Solomon polynomials, we present a generalised construction method for algebraic
GF($2^m$) codes that are applicable as LDPC codes. The construction for binary codes using idempotents has been
investigated by Shibuya and Sakaniwa~\cite{Shibuya_et_al.2003}, however, their investigation was mainly focused
on half-rate codes. In this paper, we construct some higher code-rate non binary LDPC codes with good convergence
properties. We focus on the design of short block-length
LDPC codes in view of the benefits for thin data-storage, wireless, command/control data reporting and watermarking
applications. One of the desirable features in any code construction technique is an effective method of
determining the $d_{min}$ and this feature is not present in irregular code construction methods. With our
idempotent-based method, the $d_{min}$ of a constructed code can be easily lower-bounded using the well-known
BCH bound.

The rest of the paper is organised as follows.
In Section~\ref{sec:cyclotomic_coset}, we briefly review the theory of the cyclotomic cosets, idempotents and
Mattson-Solomon polynomials. Based on the theory, we devise a generalised construction algorithm and present
an example in Section~\ref{sec:construction}.
We also outline an efficient and systematic algorithm to search for algebraic LDPC codes in
Section~\ref{sec:construction}. In Section~\ref{sec:performance}, we demonstrate the performance of the constructed
codes by means of simulation and Section~\ref{sec:conclusion} concludes this paper.

\section{Cyclotomic Cosets, Idempotents and Mattson-Solomon Polynomials}\label{sec:cyclotomic_coset}
We briefly review the theory of cyclotomic cosets, idempotents and
Mattson-Solomon polynomials to make this paper relatively
self-contained. Let us first introduce some notations that will be
used throughout this paper. Let $m$ and $m^\prime$ be positive integers
with $m|m^\prime$, so that $\text{GF}(2^m)$ is a subfield of
$\text{GF}(2^{m^\prime})$. Let $n$ be a positive odd integer and
$\text{GF}(2^{m^\prime})$ be the splitting field for $1+x^n$ over
$\text{GF}(2^m)$, so that $n|2^{m^\prime}-1$. Let
$r=(2^{m^\prime}-1)/n$, $l=(2^{m^\prime}-1)/(2^m-1)$, $\alpha$ be a generator for
$\text{GF}(2^{m^\prime})$ and $\beta$ be a generator for $\text{GF}(2^m)$, where $\beta=\alpha^l$.
Let $T_a(x)$ be the set of polynomials of degree at most $n-1$ with coefficients in
$\text{GF}(2^a)$.

\begin{definition}\label{def:ms}
If $a(x) \in T_{m^\prime}(x)$, then the finite-field transform of $a(x)$
is:
\begin{align}
A(z) &= \text{MS}\left(a(x)\right) = \sum_{j=0}^{n-1} a(\alpha^{-rj})z^j
\label{eqn:mattson_solomon}
\end{align}
where $A(z) \in T_{m^\prime}(z)$. This transform is widely known as the
Mattson-Solomon polynomial. The inverse transform is:
\begin{align}
a(x) &= \text{MS}^{-1}\left(A(z)\right) = \frac{1}{n} \sum_{i=0}^{n-1} A(\alpha^{ri})x^i
\label{eqn:inverse_mattson_solomon}
\end{align}
\end{definition}
\begin{definition}\label{def:idempotent}
Consider $e(x) \in T_m(x)$, $e(x)$ is an idempotent if the property of $e(x) = e(x)^2\text{ mod }(1+x^n)$
is satisfied. In the case of $m=1$, the property of $e(x) = e(x^2)\text{ mod }(1+x^n)$ is also satisfied.
\end{definition}

An $(n,k)$ cyclic code $\mathcal{C}$ can be described by the
generator polynomial $g(x) \in T_m(x)$ of degree $n-k$ and the
parity-check polynomials $h(x) \in T_m(x)$ of degree $k$ such that
$g(x)h(x) = 1+x^n$. It is widely known that idempotents can be
used to generate $\mathcal{C}$. Any $\text{GF}(2^m)$ cyclic code
can also be described by a unique idempotent
$e_g(x) \in T_m(x)$ which consists of a sum of primitive
idempotents. This unique idempotent is known as the generating
idempotent and, as the name implies, $g(x)$ is a divisor of this
idempotent, i.e. $e_g(x) = m(x)g(x)$, where  $m(x)$ contains the
repeated factors or non-factors of $1+x^n$.

\begin{lemma}\label{lemma:ms_idempotent}
If $e(x) \in T_m(x)$ is an idempotent, $E(z) = \text{MS}(e(x)) \in T_1(z)$.
\end{lemma}
\begin{proof} (cf.~\cite[Ch 8]{MacWilliams_et_al.1977})
Since $e(x) = e(x)^2\text{ mod }(1+x^n)$, from equation~\ref{eqn:mattson_solomon}, it
follows that $e(\alpha^{-rj}) = e(\alpha^{-rj})^2$, $\forall j \in \{0,1,\ldots,n-1\}$ 
for some integers $r$ and $l$.
Clearly, $e(\alpha^{-rj}) \in \{0,1\}$ implying that
$E(z)$ is a binary polynomial.
\end{proof}

\begin{definition}\label{def:cyclotomic_cosets}
If $s$ is a positive integer, the binary cyclotomic coset of
$s\!\!\mod n$ is:
\begin{align*}
C_s &= \left\{2^is\text{ mod } n\;|\;0 \le i \le t \right\},
\end{align*}
where we shall always assume that the subscript, $s$, is the
smallest element in the set $C_s$, and $t$ is the smallest
positive integer with the property that $2^{t+1}s = s \text{ mod }
n$. If $\mathcal{N}$ is the set consisting of the smallest
elements of all possible cyclotomic cosets then
\begin{align*}
C &= \bigcup_{s \in \mathcal{N}} C_s = \{0,1,2,\ldots,n-1\}.
\end{align*}
\end{definition}
%

\begin{lemma}\label{lemma:nonbinary_idempotent}
Let $s\in\mathcal{N}$ and let $C_{s,i}$ represents the $i$th
element of $C_s$.  Let the polynomial $e_s(x) \in T_m(x)$ be given
by
\begin{align}
e_s(x) &= \sum_{0 \le i \le |C_s|-1} e_{C_{s,i}} x^{C_{s,i}},
\end{align}
where $|C_s|$ is the number of elements in $C_s$ and $e_{C_{s,i}}$
is defined below.
\newcounter{romancount}
\begin{list}{\roman{romancount})}
{\usecounter{romancount}\setlength{\labelwidth}{0.5cm}\setlength{\leftmargin}{1cm}\setlength{\rightmargin}{0.3cm}}
\item if $m=1$,\ $e_{C_{s,i}} = 1$,
\item if $m>1$,\  $e_{C_{s,i}}$ is defined recursively as follows:
$$\begin{array}{ll} \text{for}\ i=0,
&e_{C_{s,i}}\in\{1,\beta,\beta^2,\ldots,\beta^{2^m-2}\},\\
\text{for}\ i>0,&e_{C_{s,i}}=e^2_{C_{s,i-1}}.
\end{array}$$
%
\end{list}
The polynomial so defined, $e_s(x)$, is an idempotent. We term
$e_s(x)$ a \emph{cyclotomic} idempotent.
\end{lemma}

\begin{definition}\label{def:parity_check_idempotent}
Let $\mathcal{M}\subseteq\mathcal{N}$ and let  $u(x) \in T_m(x)$
be
\begin{align}
u(x) &= \sum_{s \in \mathcal{M}} e_s(x).\label{eqn:ux}
\end{align}
Then (refer to lemma~\ref{lemma:nonbinary_idempotent}) $u(x)$ is
an idempotent and we call $u(x)$  a \emph{parity-check}
idempotent.
\end{definition}


The parity-check idempotent $u(x)$ can be used to describe the
code $\mathcal{C}$,  the parity check matrix being made up of the
$n$ cyclic shifts of the polynomial $x^{\text{deg}(u(x))} u(x^{-1})$. 

If $\left(u(x), 1+x^n\right) = h(x)$\footnote{$\left(a,b\right)$
denotes the greatest common divisor of $a$ and $b$} then, in
general, $wt(u(x))$ is much lower than
$wt(h(x))$\footnote{$wt(f(x))$ denotes the weight of polynomial
$f(x)$.}.\  Based on this observation and the fact that $u(x)$
contains all the roots of $h(x)$, we can construct cyclic codes
that have a low-density parity-check matrix.

\begin{definition}\label{def:difference_enumerator}
Let the polynomial $f(x) \in T_1(x)$. The difference enumerator of $f(x)$, 
denoted as $\mathcal{D}(f(x))$, is defined as follows:
\begin{align}
\mathcal{D}(f(x)) = f(x)f(x^{-1}) = d_0 + d_1x + \ldots + d_{n-1}x^{n-1}.\label{eqn:difference_enumerator}
\end{align}
where we assume that $\mathcal{D}(f(x))$ is a modulo $1-x^n$ polynomial with real coefficients.
\end{definition}

\begin{lemma}\label{lemma:orthogonal}
Let $m=1$ and let $d_i$ for $0 \le i \le n-1$ denote the
coefficients of $\mathcal{D}(u(x))$. If $d_i \in \{0,1\}$, $\forall
i \in \{1,2,\ldots,n-1\}$, the parity-check polynomial derived
from $u(x)$ is orthogonal on each position in the $n$-tuple.
Consequently (i) the $d_{min}$ of the resulting $\mathcal{C}$ is
$1+wt(u(x))$ and (ii) the underlying Tanner Graph has girth of at
least $6$.
\end{lemma}
\begin{proof} (i) (cf.~\cite[Theorem 10.1]{Peterson_et_al.1972})
Let a codeword $c(x)=c_0 + c_1x + \ldots + c_{n-1}x^{n-1}$ and $c(x) \in T_1(x)$.
For each non zero bit position $c_j$ of $c(x)$ where $j \in \{0,1,\ldots,n-1\}$,
there are $wt(u(x))$ parity-check equations orthogonal to position $c_j$. Each of the parity-check
equation must check another non zero bit $c_l$ $l \ne j$ so that the equation is satisfied.
Clearly, $wt(c(x))$ must equal to $1+wt(u(x))$ and this is the
minimum weight of all codewords.
(ii) The direct consequence of having orthogonal parity-check equation is the absence of cycles of length $4$
in the Tanner Graphs. It can be shown that there exists three integers $a$, $b$ and $c$, such that
$2(b-a) \equiv (c-b)$ for $a < b < c$. If these three integers are associated to the variable nodes in the
Tanner Graphs, a cycle of length $6$ can be formed between these variable nodes and some check nodes.
\end{proof}

From Lemma~\ref{lemma:orthogonal} we can deduce that $u(x)$ is the parity-check polynomial for
One-Step Majority-Logic Decodable codes if $d_i \in \{0,1\}$, $\forall i \in \{1,2,\ldots,n-1\}$ or
the parity-check polynomial for Difference-Set Cyclic codes if $d_i = 1$, $\forall i \in \{1,2,\ldots,n-1\}$.

\begin{lemma}\label{lemma:nonbinary_dmin_bound}
For the non binary GF($2^m$) cyclic codes, the $d_{min}$ is bounded by:
\begin{align*}
d_0 < d_{min} \le \text{min}\left(wt(g(x)), 1+wt(u(x))\right)
\end{align*}
where $d_0$ denotes the maximum run of consecutive ones in $U(z)$ taken cyclically modulo $n$.
\end{lemma}
\begin{proof}
The lower-bound of the $d_{min}$ of a cyclic code, BCH bound is determined from the number of consecutive
roots of $e_g(x)$ and from lemma~\ref{lemma:ms_idempotent}, it is equivalent to the run of consecutive ones
in $U(z)$.
\end{proof}


\section{Construction Algorithm for the Codes}\label{sec:construction}
Based on the mathematical theories outlined above, we devise an algorithm to construct
GF($2^m$) $\mathcal{C}$ which are applicable for iterative decoding.
The construction algorithm can be described in the following procedures:
\begin{enumerate}
\item Given the integers $m$ and $n$, find the splitting field ($\text{GF}(2^{m^\prime})$) of $1+x^n$ over $\text{GF}(2^m)$.
We can only construct $\text{GF}(2^m)$ cyclic codes of length $n$ if and only if the condition of $m|m^\prime$ is
satisfied.
\item Generate the cyclotomic cosets modulo $2^{m^\prime}-1$ and denote it $C^\prime$.
\item Derive a polynomial $p(x)$ from $C^\prime$. Let $s \in \mathcal{N}$ be the
smallest positive integer such that $|C^\prime_s| = m$. The polynomial $p(x)$ is the minimal polynomial
of $\alpha^s$:
\begin{align}
p(x) = \prod_{0 \le i < m} \left(x+\alpha^{C^\prime_{s,i}}\right)
\end{align}
Construct all elements of GF($2^m$) using $p(x)$ as the primitive polynomial.
\item Let $C$ be the cyclotomic cosets modulo $n$ and $\mathcal{N}$ be a set containing the smallest number
in each coset of $C$. Assume that there exists a non empty set $\mathcal{M} \subset \mathcal{N}$ and following
definition~\ref{def:parity_check_idempotent}, construct the parity-check idempotent $u(x)$.
The coefficients of $u(x)$ can be assigned following lemma~\ref{lemma:nonbinary_idempotent}.
\item Generate the parity-check matrix of $\mathcal{C}$ using the $n$ cyclic shifts of $x^{\text{deg}(u(x))} u(x^{-1})$.
\item Compute $r$ and $l$, then take the Mattson-Solomon polynomial of $u(x)$ to produce $U(z)$.
Obtain the code dimension and the lower-bound of the $d_{min}$ from $U(z)$.
\end{enumerate}
Note that care should be taken to ensure that there is no common factor between $n$ and all of the exponents of
$u(x)$, apart from unity, in order to avoid a degenerate code.

\begin{table*}[!b]
\renewcommand{\arraystretch}{1.3}
\caption{\label{tbl:code-parameter}Code examples}
\centering
\begin{tabular}{c|p{2.5in}|c|c|c|c}\hline
$\mathcal{C}$ & \multicolumn{1}{c|}{$u(x)$} & $d_{min}$ & $d_b^\dagger$ & Comment & SPB$^\ddagger$\\\hline
$\text{GF}(4)$ $(51,29)$ & $\beta^2 x^3 + \beta x^6 + \beta^2 x^{12}, x^{17} + \beta x^{24} + \beta x^{27} + x^{34} +
\beta^2 x^{39} + \beta x^{45} + \beta^2 x^{48}$ & $5$ & $10$ & $m=2$, $m^\prime=8$, $r=5$ and $l=85$ & $0.25$dB\\\hline
$\text{GF}(4)(255,175)$ & 
$\beta x^7 + \beta^2x^{14} + \beta x^{28} + \beta^2x^{56} + x^{111} + \beta x^{112} +
x^{123} + \beta^2 x^{131} + x^{183} + x^{189} + \beta x^{193} + x^{219} + x^{222} + \beta^2x^{224} + x^{237} +
x^{246}$ & 
$\ge 17$ & $20$ & $m=2$, $m^\prime=8$, $r=1$ and $l=85$ & $0.36$dB\\\hline
$\text{GF}(4)(273,191)$ & 
$\beta^2x^{23} + \beta x^{37} + \beta x^{46} + \beta^2x^{74} + \beta x^{91} + \beta^2x^{92} + \beta^2x^{95} +
\beta^2x^{107} + x^{117} + \beta x^{148} + \beta^2x^{155} + \beta^2x^{182} + \beta x^{184} + \beta x^{190} +  
x^{195} + \beta x^{214} + x^{234}$ & 
$\ge 18$ & $20$ &
$m=2$, $m^\prime=12$, $r=15$ and $l=1365$ & $0.4$dB\\\hline
$\text{GF}(8)(63,40)$ & $1 + \beta^5x^9 + \beta x^{13} + \beta^3x^{18} + \beta^2x^{19} + \beta^2x^{26} +
\beta^6x^{36} + \beta^4x^{38} + \beta x^{41} + \beta^4x^{52}$ & $\ge 6$ & $10$ &
$m=3$, $m^\prime=6$, $r=1$ and $l=9$ & $0.3$dB\\\hline
$\text{GF}(8)$ $(63,43)$ & $\beta^2 x^9 + \beta^3 x^{11} + \beta^4 x^{18} + x^{21} + \beta^6 x^{22} +
\beta^3 x^{25} + x^{27} + \beta x^{36} + \beta^5 x^{37} + x^{42} + \beta^5 x^{44} + x^{45} + \beta^6 x^{50} + x^{54}$ &
$\ge 8$ & $12$ & $m=3$, $m^\prime=6$, $r=1$ and $l=9$ & $0.45$dB\\\hline
$\text{GF}(8)(91,63)$ & 
$\beta^6x + \beta^5x^2 + \beta^3x^4 + \beta^6x^8 + \beta x^{13} + \beta^5x^{16} + \beta^5x^{23} + \beta^2x^{26} + \beta^3x^{32} +
\beta^5x^{37} + \beta^3x^{46} + \beta^4x^{52} + \beta^6x^{57} +  
\beta^6x^{64} + \beta^3x^{74}$ & 
$\ge 8$ & $10$ & $m=3$, $m^\prime=12$, $r=45$ and $l=585$ & $0.35$dB\\\hline
$\text{GF}(32)(31,20)$ & $1 + \beta^{28}x^5 + \beta^7x^9 + \beta^{25}x^{10} + x^{11} + x^{13} + \beta^{14}x^{18} +
\beta^{19}x^{20} + x^{21} + x^{22} + x^{26}$ & $\ge 7$ & $12$ & $m=5$, $m^\prime=5$, $r=l$ and $l=1$ & $0.4$dB\\\hline
$\text{GF}(32)(31,21)$ & $\beta^{23}x^5 + \beta^{29}x^9 + \beta^{15}x^{10} + \beta x^{11} + \beta^4x^{13} +
\beta^{27}x^{18} + \beta^{30}x^{20} + \beta^{16}x^{21} + \beta^2x^{22} + \beta^8x^{26}$ & $\ge 4$ & $8$ &
$m=5$, $m^\prime=5$, $r=1$ and $l=1$ & $0.25$dB\\\hline
\end{tabular}
\begin{flushleft}
{\scriptsize\textsuperscript{$^\dagger$}The code minimum distance in binary level.\\}
{\scriptsize\textsuperscript{$^\ddagger$}Distance to the sphere-packing-bound constrained for binary transmission.}
\end{flushleft}
\end{table*}

\begin{example}\label{ex:code-construction}
Let us assume that we want to construct a $\text{GF}(64)$ $n=21$ cyclic idempotent code. The splitting field for
$1+x^{21}$ over $\text{GF}(64)$ is $\text{GF}(64)$ and this implies that $m=m^\prime=6$, $r=3$ and $l=1$. Let $C$ and
$C^\prime$ denote the cyclotomic cosets modulo $n$ and $2^{m^\prime}-1$ respectively. $|C^\prime_1| = 6$ and therefore the
primitive polynomial $p(x)$ has roots of $\alpha^j$, $\forall j \in C^\prime_1$, i.e. $p(x)=1+x+x^6$.
By letting $1+\beta+\beta^6=0$, all of the elements of $\text{GF}(64)$
can be defined. If we let $u(x)$ be the parity-check idempotent generated by the sum of the cyclotomic idempotents
defined by $C_s$ where $s\in\{\mathcal{M}:5,7,9\}$ and $e_{C_{s,0}}$, $\forall s \in\mathcal{M}$ be
$\beta^{23}$, $1$ and $1$ respectively, $u(x) = \beta^{23}x^5 + x^7 + x^9 + \beta^{46}x^{10} +
\beta^{43}x^{13} + x^{14} + x^{15} + \beta^{53}x^{17} + x^{18} + \beta^{58}x^{19} + \beta^{29}x^{20}$ and its
Mattson-Solomon polynomial $U(z)$ tells us that it is $\text{GF}(64)(21,15)$ cyclic code with $d_{min} >= 5$.
\end{example}

A systematic algorithm has been developed to sum up all combinations of the cyclotomic idempotents
to search for all possible $\text{GF}(2^m)$ cyclic codes of a given length.
The search algorithm is targeted on the following key parameters:
\begin{enumerate}
\item Sparseness of the resulting parity-check matrix. Since the parity-check matrix of $\mathcal{C}$ is directly
derived from $u(x)$ which consists of the sum of the cyclotomic idempotents, we are only
interested in low-weight cyclotomic idempotents. Let us define $W_{max}$ as the maximum $wt(u(x))$ then the search algorithm
will only choose the cyclotomic idempotents whose sum has total weight less than or equal to $W_{max}$.
\item High code-rate. The number of roots of $u(x)$ which are also roots of unity define the dimension of
$\mathcal{C}$ and let us define $k_{min}$ as the minimum information length of $\mathcal{C}$. We are only interested
in the sum of the cyclotomic idempotents whose Mattson-Solomon polynomial has at least $k_{min}$ zeros.
\item High $d_{min}$. Let us define $d$ as the minimum value of the $d_{min}$ of $\mathcal{C}$. The sum of the
cyclotomic idempotents should have at least $d-1$ consecutive powers of $\beta$ which are roots of unity
but not roots of $u(x)$.
\end{enumerate}
The search algorithm can be relaxed to allow the existence of cycles of length 4 in the resulting
parity-check matrix of $\mathcal{C}$. The condition of cycles-of-length-4 is not crucial as we will show later that
there are codes that have good convergence properties when decoded using iterative decoder. Clearly, by eliminating
the cycles-of-length-4 constraint, we can construct more codes.

Following definitions~\ref{def:ms} and \ref{def:parity_check_idempotent}:
\begin{align*}
U(z) &= \text{MS}\left(\sum_{s\in\mathcal{M}} e_s(x) \right) = \sum_{s\in\mathcal{M}} E_s(z)
\end{align*}
and hence it is possible to maximise the run of the consecutive ones in $U(z)$ if the coefficients of $e_s(x)$
are aligned appropriately. It is therefore important that all possible non zero values of $e_{C_{s,0}}$,
$\forall s\in\mathcal{M}$ are included in the search in order to guarantee that we can obtain codes with the highest
possible $d_{min}$ or at least to obtain a better estimate of the $d_{min}$.

\section{Code Performance}\label{sec:performance}
\begin{figure}[hbt]
\centering
\includegraphics[width=2.25in,angle=270]{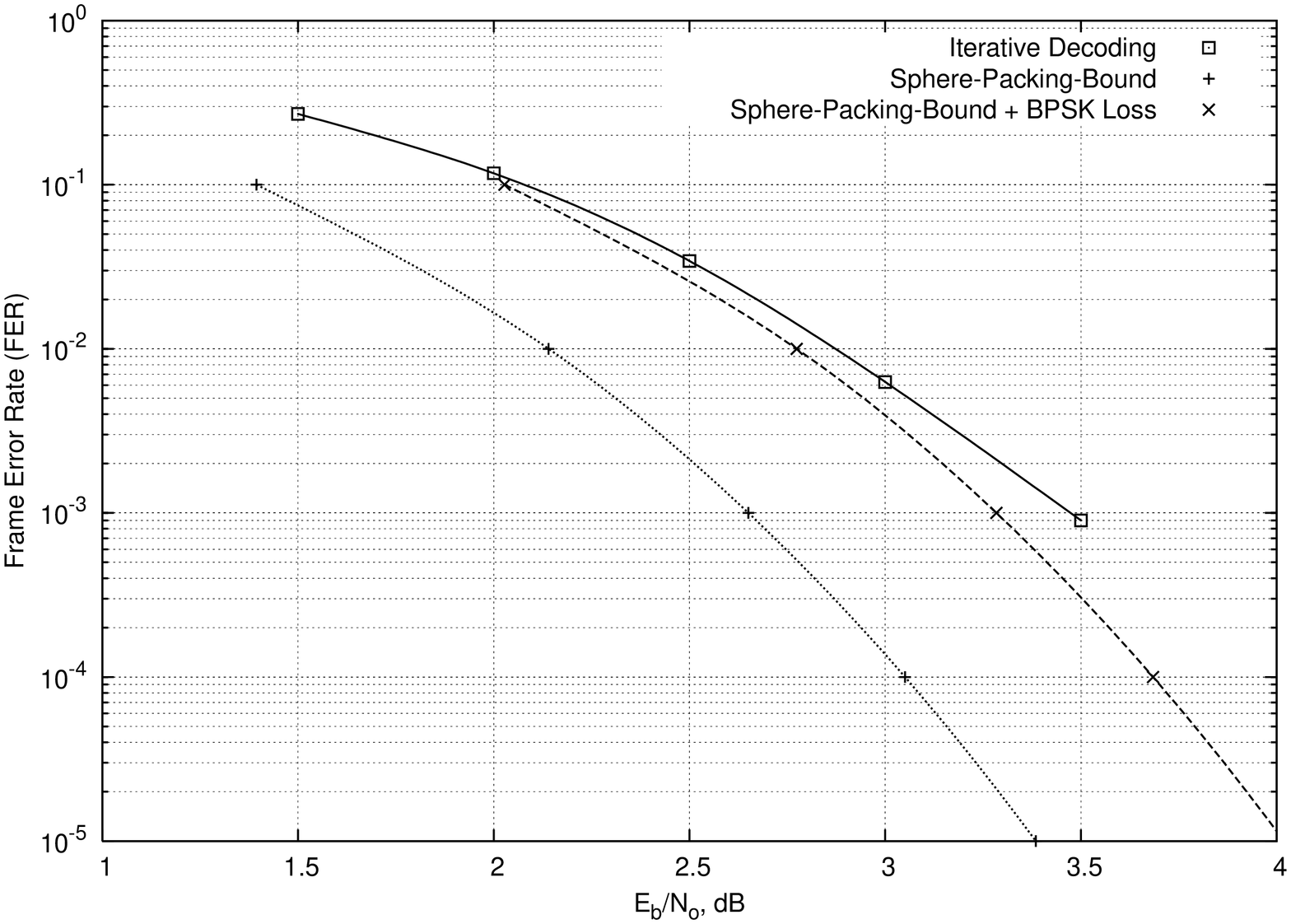}
\caption{\label{fig:fer-gf64-21-15}Frame error performance of the $\text{GF}(2^6)(21,15)$ cyclic LDPC code}
\end{figure}

As an example of the performance attainable from an iterative decoder, computer simulations have been carried out
for several $\text{GF}(2^m)$ cyclic LDPC codes. We assume BPSK signalling and the iterative decoder
used is the modified belief-propagation decoder which approximates the performance of a maximum-likelihood
decoder~\cite{Tjhai_et_al.patent},\cite{Papagiannis_et_al.isit2005}. 
The frame-error-rate (FER) performance of the $\text{GF}(2^6)(21,15)$ cyclic LDPC code is shown in Fig.~\ref{fig:fer-gf64-21-15}
and is compared with the sphere-packing-bound~\cite{Shannon.1959},\cite{Dolinar_et_al.1998} for binary codes of length
$126$ bits offset by the binary transmission loss\footnote{In the rest of this paper, we assume that the sphere-packing-bound
has been offset by the information theoretical loss associated with binary transmission.}. 
We can see that the performance of the code is within $0.2$dB away from this bound at $10^{-3}$ FER.
The binary level minimum-distance of this $\text{GF}(64)(21,15)$ cyclic LDPC code is $9$.
\begin{figure}[hbt]
\centering
\includegraphics[width=2.25in,angle=270]{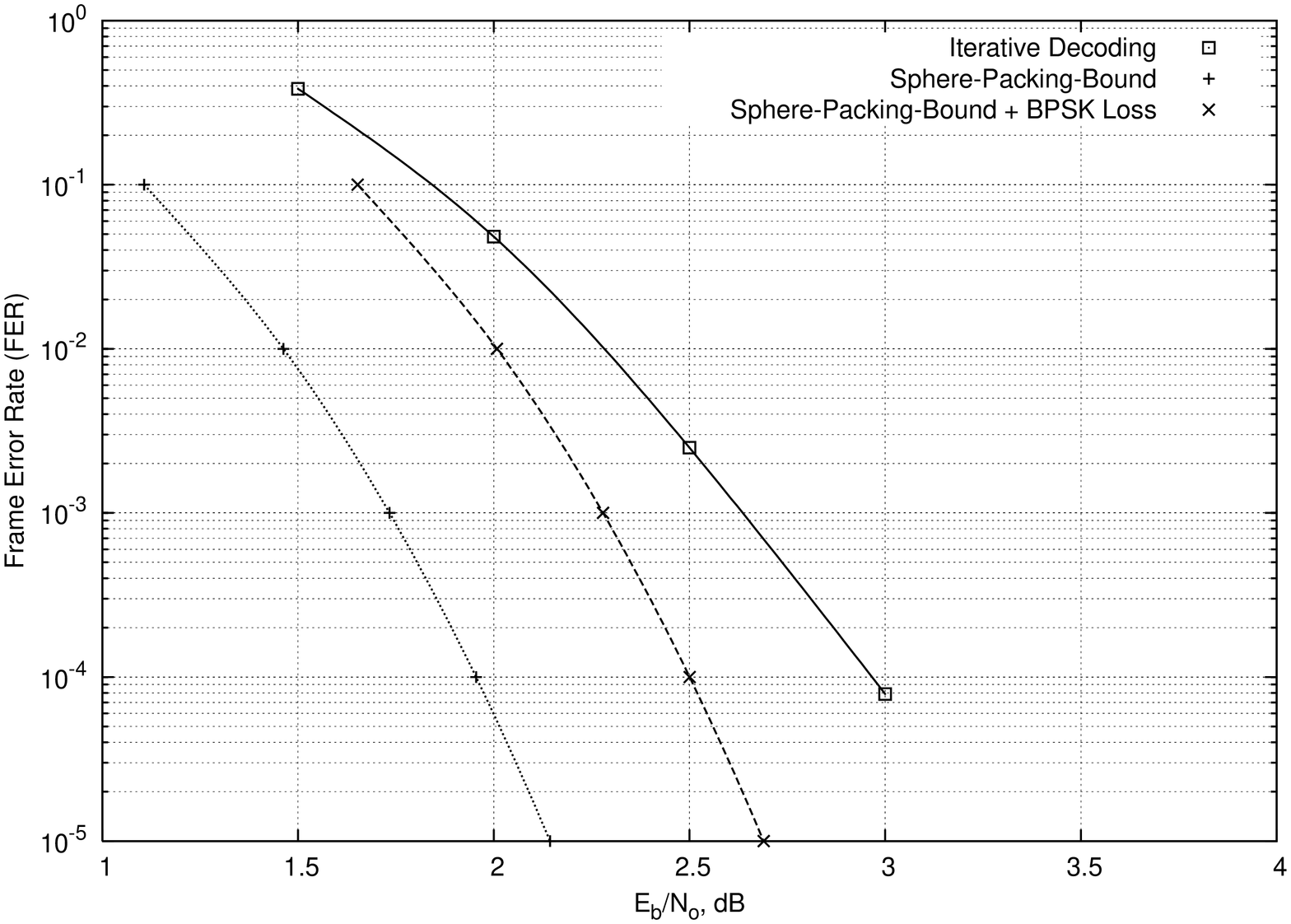}
\caption{\label{fig:fer-gf4-255-175}Frame error performance of the $\text{GF}(2^2)(255,175)$ cyclic LDPC code}
\end{figure}
\begin{figure}[hbt]
\centering
\includegraphics[width=2.25in,angle=270]{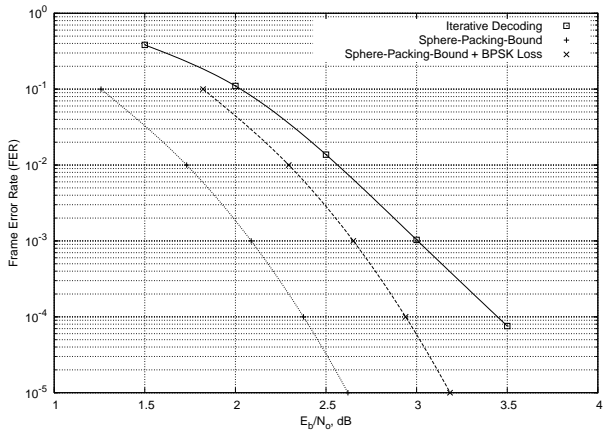}
\caption{\label{fig:fer-gf8-91-63}Frame error performance of the $\text{GF}(2^3)(91,63)$ cyclic LDPC code}
\end{figure}

Fig.~\ref{fig:fer-gf4-255-175} shows the FER curve of the $\text{GF}(2^2)(255,175)$ cyclic LDPC code which
is equivalent to $(510,350)$ binary code. At $10^{-3}$ FER, the performance of this code is approximately $0.36$dB away
from the sphere-packing-bound of length $510$ bits. While both of the codes mentioned above are free from cycles
of length 4, good convergence codes exist even if they have cycles of length 4 in the underlying Tanner Graph. One such
example is the $\text{GF}(2^3)(91,63)$ cyclic code whose FER performance is shown in Fig.~\ref{fig:fer-gf8-91-63}.
At $10^{-3}$ FER, the code performs around $0.35$dB away from the sphere-packing-bound of length $273$ bits. The parameters
of the codes in Fig.~\ref{fig:fer-gf4-255-175} and \ref{fig:fer-gf8-91-63} are available in
Table~\ref{tbl:code-parameter}. Some other examples of the non binary $\text{GF}(2^m)$ cyclic LDPC codes with
their parameters and distance from the sphere-packing-bound are also shown in Table~\ref{tbl:code-parameter}.

\section{Conclusions}\label{sec:conclusion}
An algebraic construction technique for GF($2^m$) ($m>0$) LDPC codes based on summing the cyclotomic idempotents
to define the parity-check polynomial is able to produce a large number of cyclic codes. The fact that we consider
step-by-step summation of the cyclotomic idempotents, we are able to control the sparseness of the resulting parity-check
matrix. The lower-bound of the $d_{min}$
and the dimension of the codes can be easily determined from the Mattson-Solomon polynomial of the resulting idempotent.
For GF($2$) case where the parity-check polynomials are orthogonal on each bit position, we can even determine the true
$d_{min}$ of the codes regardless of the code length. In fact, this special class of binary cyclic codes are the
Difference-Set Cyclic and the One-Step Majority-Logic Decodable codes which can be easily constructed using our method.
For non-binary cases, if the constructed code has low $d_{min}$, we can concatenate this code with an inner binary code
to trade improvement in $d_{min}$ with loss in code-rate.

Simulation results have shown that these codes can converge well under iterative decoding and their performance is very
close to the sphere-packing-bound of binary codes for the same code length and rate. The excellent performance of these
codes coupled with their low-complexity encoder offers an attractive coding scheme for applications that required short
block-lengths such as thin data-storage, wireless, command/control data reporting and watermarking.

\section*{Acknowledgement}
This research is partially funded by the UK Overseas Research Students Award Scheme.

\balance


\begin{thebibliography}{10}

\bibitem{Richardson_et_al.2001b}
T.~J. Richardson, M.~A. Shokrollahi, and R.~L. Urbanke, ``{Design of
  Capacity-Approaching Irregular Low-Density Parity-Check Codes},'' {\em IEEE
  Trans. Inform. Theory}, vol.~47, pp.~619--637, Feb. 2001.

\bibitem{Chung_et_al.2001b}
S.~Y. Chung, G.~D. Forney{, Jr.}, T.~J. Richardson, and R.~L. Urbanke, ``{On
  the Design of Low-Density Parity Check Codes within 0.0045 dB of the Shannon
  Limit},'' {\em {IEEE Comm. Letters}}, vol.~3, pp.~58--60, Feb. 2001.

\bibitem{Chung_et_al.2001a}
S.~Y. Chung, T.~J. Richardson, and R.~L. Urbanke, ``{Analysis of Sum-Product
  Decoding of Low-Density Parity-Check Codes Using a Gaussian Approximation},''
  {\em IEEE Trans. Inform. Theory}, vol.~47, pp.~657--670, Feb. 2001.

\bibitem{Richardson_et_al.2001a}
T.~J. Richardson and R.~L. Urbanke, ``{The Capacity of Low-Density Parity-Check
  Codes Under Message-Passing Decoding},'' {\em IEEE Trans. Inform. Theory},
  vol.~47, pp.~599--618, Feb. 2001.

\bibitem{Campello_et_al.2001}
J.~Campello and D.~S. Modha, ``{Extended Bit-Filling and LDPC Code Design},''
  {\em {Proc. of the IEEE Globecom Conf.}}, pp.~25--29, Nov. 2001.

\bibitem{Hu_et_al.2002}
X.~Y. Hu, E.~Eleftheriou, and D.~M. Arnold, ``{Irregular Progressive
  Edge-Growth Tanner Graphs},'' {\em {Proc. of IEEE Intl. Symp. Inform. Theory
  (ISIT), Lausanne, Switzerland}}, July 2002.

\bibitem{Ramamoorthy_et_al.2004b}
A.~Ramamoorthy and R.~D. Wesel, ``{Construction of Short Block Length Irregular
  Low-Density Parity-Check Codes},'' {\em IEEE Int. Conf. Comm.}, June 2004.

\bibitem{Kou_et_al.2001}
Y.~Kou, S.~Lin, and M.~Fossorier, ``Low density parity check codes based on
  finite geometries: A rediscovery and new results,'' {\em IEEE Trans. Inform.
  Theory}, vol.~47, pp.~2711--2736, Nov. 2001.

\bibitem{Davey_et_al.1998}
M.~C. Davey and D.~J.~C. MacKay, ``{Low-Density Parity-Check Codes over
  GF(q)},'' {\em IEEE Comm. Letters}, vol.~2, pp.~165--167, June 1998.

\bibitem{Hu_et_al.2005}
X.~Y. Hu, E.~Eleftheriou, and D.~M. Arnold, ``{Regular and Irregular
  Progressive Edge-Growth Tanner Graphs},'' {\em IEEE Trans. Inform. Theory},
  vol.~51, pp.~386--398, Jan. 2005.

\bibitem{MacWilliams_et_al.1977}
F.~J. MacWilliams and N.~J.~A. Sloane, {\em {The Theory of Error-Correcting
  Codes}}.
\newblock {North-Holland}, 1977.
\newblock ISBN 0 444 85193 3.

\bibitem{MacWilliams.1979}
F.~J. MacWilliams, ``A Table of Primitive Binary Idempotents of Odd Length $n$,
  $7\,\le\,n\,\le\,511$,'' {\em IEEE Trans. Inform. Theory}, vol.~IT-25,
  pp.~118--123, Jan. 1979.

\bibitem{Shibuya_et_al.2003}
T.~Shibuya and K.~Sakaniwa, ``Construction of Cyclic Codes Suitable for Iterative
  Decoding via Generating Idempotents,'' {\em {IEICE Trans. Fundamentals}},
  vol.~E86-A, no.~4, 2003.

\bibitem{Peterson_et_al.1972}
W.~Peterson and E.~J. Weldon{, Jr.}, {\em {Error-Correcting Codes}}.
\newblock {MIT Press.}, 1972.

\bibitem{Tjhai_et_al.patent}
C.~J. Tjhai, E.~Papagiannis, M.~Tomlinson, M.~A. Ambroze, and M.~Z. Ahmed,
  ``Improved iterative decoder for {LDPC} codes with performance approximating
  to a maximum likelihood decoder.'' {UK Patent Application 0409306.8}, Apr.
  2004.

\bibitem{Papagiannis_et_al.isit2005}
E.~Papagiannis, M.~Ambroze, and M.~Tomlinson, ``{Improved Decoding of
  Low-Density Parity-Check Codes with Low, Linearly Increased Added
  Complexity}.'' Submitted to 4\textsuperscript{th} IASTED International Conference
  on Communication Systems and Networks, 2005.

\bibitem{Shannon.1959}
C.~E. Shannon, ``Probability of error for optimal codes in a gaussian
  channel,'' {\em Bell Syst. Tech. J.}, vol.~38, pp.~611--656, May 1959.

\bibitem{Dolinar_et_al.1998}
S.~Dolinar, D.~Divsalar, and F.~Pollara, ``Code performance as a function of
  block size,'' {\em {TMO Progress Report}}, pp.~42--133, May 1998.

\end{thebibliography}
\end{document}